\documentclass{moriond}
\usepackage{wrapfig}

\providecommand{\abstracts}[1]{%
  \begin{center}%
  \begin{minipage}{0.88\textwidth}%
  \small #1%
  \end{minipage}%
  \end{center}%
  \vspace{0.4cm}%
}

\def\Journal#1#2#3#4{{#1} {\bf #2}, #3 (#4)}

\def\NIMA{{\em Nucl. Instrum. Methods} A}

\def\PRL{\em Phys. Rev. Lett.}
\def\PRD{{\em Phys. Rev.} D}

\def\APJ{{\em Astrophys. J.}}
\def\APJL{{\em Astrophys. J. Lett.}}
\def\ASP{{\em Astropart. Phys.}}
\def\PTEP{{\em Prog. Theor. Exp. Phys.}}
\def\EPJ{{\em EPJ Web Conf.}}

\def\be{\begin{equation}}
\def\ee{\end{equation}}
\def\bea{\begin{eqnarray}}
\def\eea{\end{eqnarray}}

\providecommand{\Photo}{}

\begin{document}
\vspace*{4cm}
\title{Recent Findings from the Telescope Array Experiment}

\author{Jihyun Kim$^{1,2,3}$ on behalf of the Telescope Array Collaboration}

\address{$1$Graduate School of Science, Osaka Metropolitan University, Sugimoto, Sumiyoshi, Osaka, Japan
\\
$^2$Nambu Yoichiro Institute of Theoretical and Experimental Physics, Osaka Metropolitan University, Sugimoto, Sumiyoshi, Osaka, Japan
\\
$^3$High Energy Astrophysics Institute and Department of Physics and Astronomy, University of Utah, Salt Lake City, Utah, USA}

\maketitle
\abstracts{
The Telescope Array (TA) is the largest ultra-high energy cosmic ray (UHECR) observatory in the Northern Hemisphere. Together with its extensions, TA Low Energy (TALE), TALE infill, and the TA$\times$4 array, it measures extensive air showers (EAS) initiated by UHECRs across an energy range spanning from $10^{15}$ eV to beyond $10^{20}$ eV. All components of the experiment employ a hybrid detection approach, combining plastic scintillator arrays that sample the EAS footprint at ground level with telescopes that record fluorescence and Cherenkov light from shower development in the atmosphere. The ongoing construction of TA$\times$4 will significantly increase statistics at the highest energies by expanding the surface detector area by a factor of four. In addition, the recently deployed TALE infill array further lowers the hybrid energy threshold of TALE. This presentation summarizes the current status of the TA experiment and highlights recent findings on the energy spectrum, mass composition, and anisotropy.
}

\section{Introduction}

The origin and nature of ultra-high-energy cosmic rays (UHECRs), charged particles arriving at Earth with energies above $10^{18}$~eV, remain among the central open questions in astroparticle physics. The UHECR energy spectrum exhibits several prominent structures, including the knee, low-energy ankle, second knee, ankle, shoulder feature, and the cutoff at the highest energies. These features reflect the properties of cosmic ray sources, propagation effects, and the transition between different source populations. Mass composition and anisotropy in the arrival direction distribution provide complementary information, since the charge of the primary particle determines its magnetic deflection and thus affects the relation between arrival directions and candidate sources. Together, the energy spectrum, mass composition, and anisotropy are the key observables for understanding the origin and nature of UHECRs.

\begin{wrapfigure}{r}{0.5\textwidth}
\centering
\includegraphics[width=0.48\textwidth]{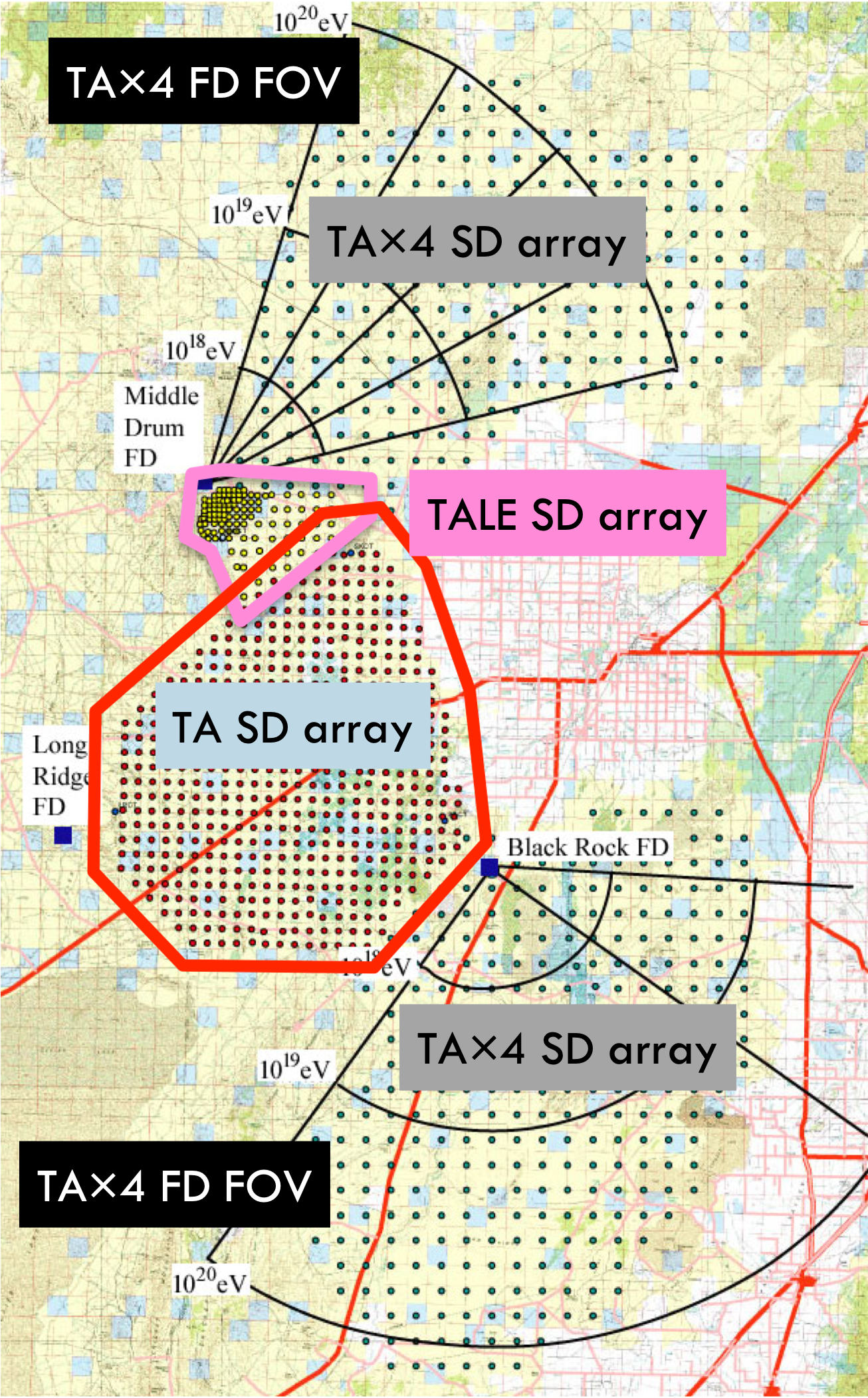}
\caption{\textbf{Map of the Telescope Array (TA) experiment together with the extensions.} The red border line indicates the TA SD array, and the pink one represents the low energy extension, TALE. The fourfold expansion of TA, TA$\times$4, is located north and south of the main TA array. All the TA projects employ a hybrid technique to observe extensive air showers induced by primary UHE particles striking the Earth's atmosphere.}
\label{fig:layout}
\end{wrapfigure}

The Telescope Array (TA) experiment is a hybrid UHECR observatory in the Northern Hemisphere located near Delta, Utah, USA, at an altitude of approximately 1400~m above sea level. Figure~\ref{fig:layout} shows the map of the experiment together with the extensions. The original TA surface detector (SD) array consists of 507 plastic scintillation counters deployed on a 1.2~km square grid and covering an area of about 700~km$^2$~\cite{TelescopeArray:2012uws}. The SD array is observed by fluorescence detectors (FDs) at three sites, Middle Drum, Long Ridge, and Black Rock Mesa, with a total of 38 telescopes for hybrid measurements~\cite{TelescopeArray:2012vhh,Tokuno:2012mi}. The SD array samples the lateral distribution and timing structure of shower particles at ground level with a high duty cycle, while the FD technique provides calorimetric measurements of the shower energy and of the atmospheric depth of the shower maximum, $X_{\max}$, which is the primary observable used in mass composition studies.

The TA Low-energy Extension (TALE) extends the energy reach of the experiment down to about $10^{15}$~eV, enabling studies of the Galactic-to-extragalactic transition region~\cite{TelescopeArray:2018bya,Komae:2023fsb}. TALE includes 10 high-elevation FD telescopes at the Middle Drum site together with an SD array of 80 counters spaced at 400--600~m. A newer TALE infill array, consisting of 50 counters with nearly 100~m spacing, has also been deployed to further lower the hybrid threshold and improve measurements in the PeV--EeV region~\cite{Kawachi:2025bbk}. At the highest energies, the TA$\times$4 project expands the original TA aperture by a factor of four, with a target coverage of $\sim$3000~km$^2$~\cite{TelescopeArray:2021dri}. This expansion includes 500 additional SDs on a 2.08~km grid and 12 new FD telescopes, and is intended to improve the statistics for source searches, composition measurements, and studies of spectral features in the highest-energy regime.

Beyond its primary goals, TA also enables studies of hadronic interactions at energies beyond those accessible to terrestrial accelerators, including measurements of the proton--air cross section and the muon content of extensive air showers~\cite{Abbasi2020_ProtonAirCrossSection_MD,Abbasi2020_ProtonAirCrossSection_BRM_LR,Abbasi2018_TA_MuonStudy}. TA also contributes to multi-messenger and exotic-particle searches, including UHE photons, neutrinos, and macroscopic dark matter~\cite{Abbasi2019_TA_DiffusePhotonFlux,Abbasi2020_TA_NeutrinoSearch,Shinozaki2023_DIMS_TA}. In addition, TA supports interdisciplinary studies of thunderstorms, downward terrestrial gamma-ray flashes, and meteoroid imaging~\cite{Belz2020_TA_DownwardTGFOrigin,Shinozaki2023_DIMS_TA}. The TA sites also serve as testbeds for the development and cross-calibration of next-generation UHECR detectors~\cite{JEMEUSO2015_EUSOTA_GroundTests,Malacari2020_FAST_Prototypes,Tameda2023_CRAFFT,Mayotte2023_AugerAtTA}.

\section{Recent Findings from the Telescope Array Experiment}
\label{sec:results}
The following subsections summarize recent TA results on the energy spectrum, mass composition, and anisotropy. Spectrum measurements identify features such as the ankle, shoulder, and high-energy cutoff. Composition measurements based on $X_{\max}$ probe primary-mass evolution in the PeV--EeV region and extend to the highest energies with TA$\times$4 hybrid events. Arrival-direction studies examine correlations with nearby large-scale structure and intermediate-scale excesses in the Northern sky.

\subsection{Energy Spectrum}
\label{subsec:spectrum}

\begin{figure}[tbp]
\centering
\includegraphics[width=0.85\columnwidth]{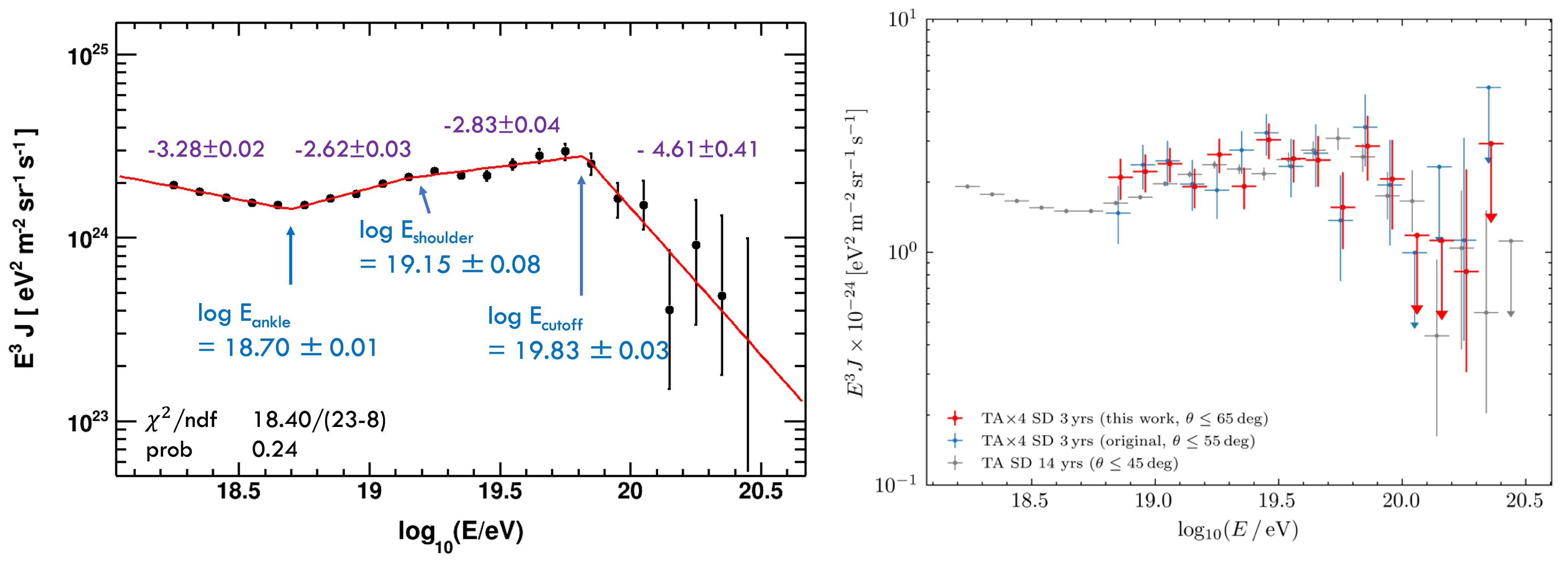}
\caption{\textbf{TA and TA$\times$4 SD energy spectra.} \textbf{Left}: TA SD spectrum plotted with a broken power law fit (red), showing the ankle, shoulder, and high-energy cutoff. The spectral features are given on the plot. \textbf{Right}: Comparison of the 3-year TA$\times$4 SD spectra from the updated $\theta \leq 65^\circ$ analysis (red) and the original $\theta \leq 55^\circ$ analysis (blue), together with the 14-year TA SD result (gray). The consistency with the TA spectrum and the earlier TA$\times$4 result confirms the robustness of the measurement, while the wider zenith-angle range improves exposure at the highest energies.}
\label{fig:ta_tax4_spec}
\end{figure}

\begin{figure}[tbp]
\centering
\includegraphics[width=0.85\columnwidth]{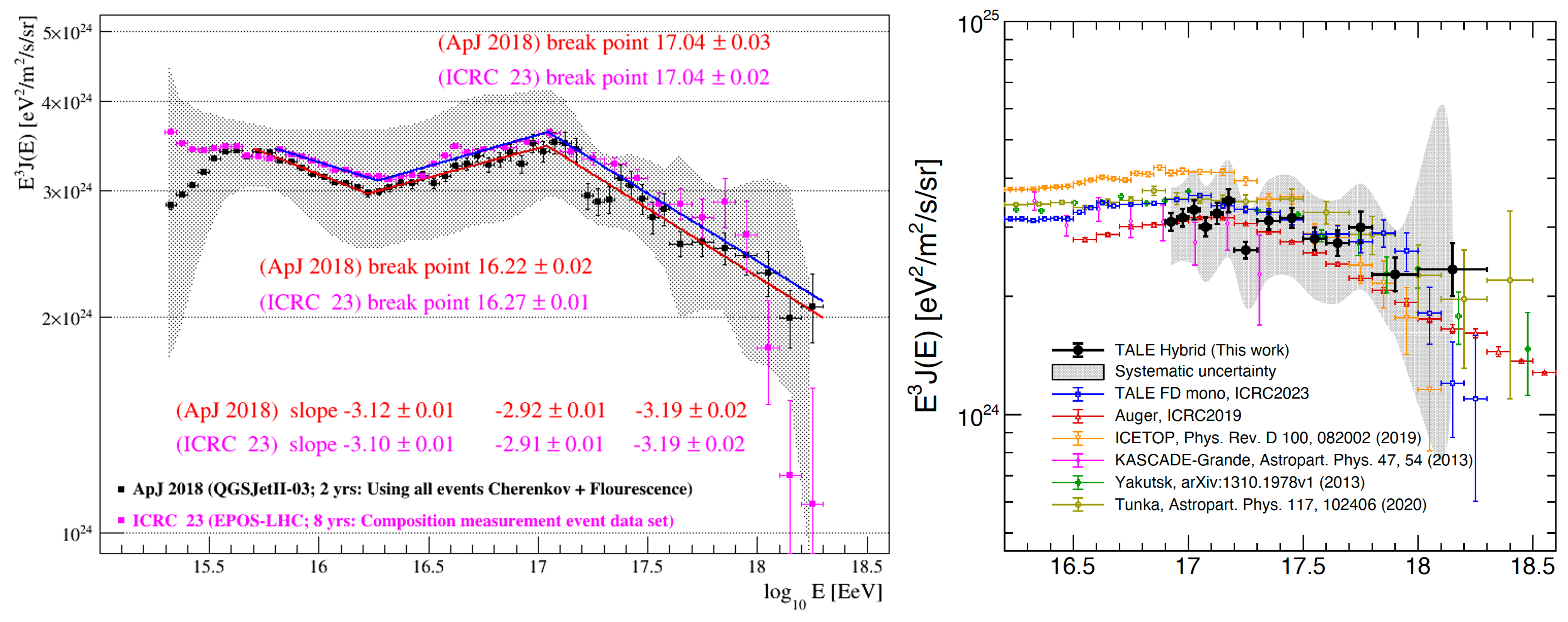}
\caption{\textbf{TALE energy specta.} \textbf{Left}: Comparison of the earlier 2-year and updated 8-year TALE FD monocular spectra with broken power law fits. Different hadronic models are used. \textbf{Right}: TALE hybrid spectrum with the TALE FD monocular result and selected results from other experiments. The measurements are consistent with previous observations in the PeV--EeV transition region.}
\label{fig:tale_spec}
\end{figure}

\begin{figure}[tbp]
\centering
\includegraphics[width=0.7\columnwidth]{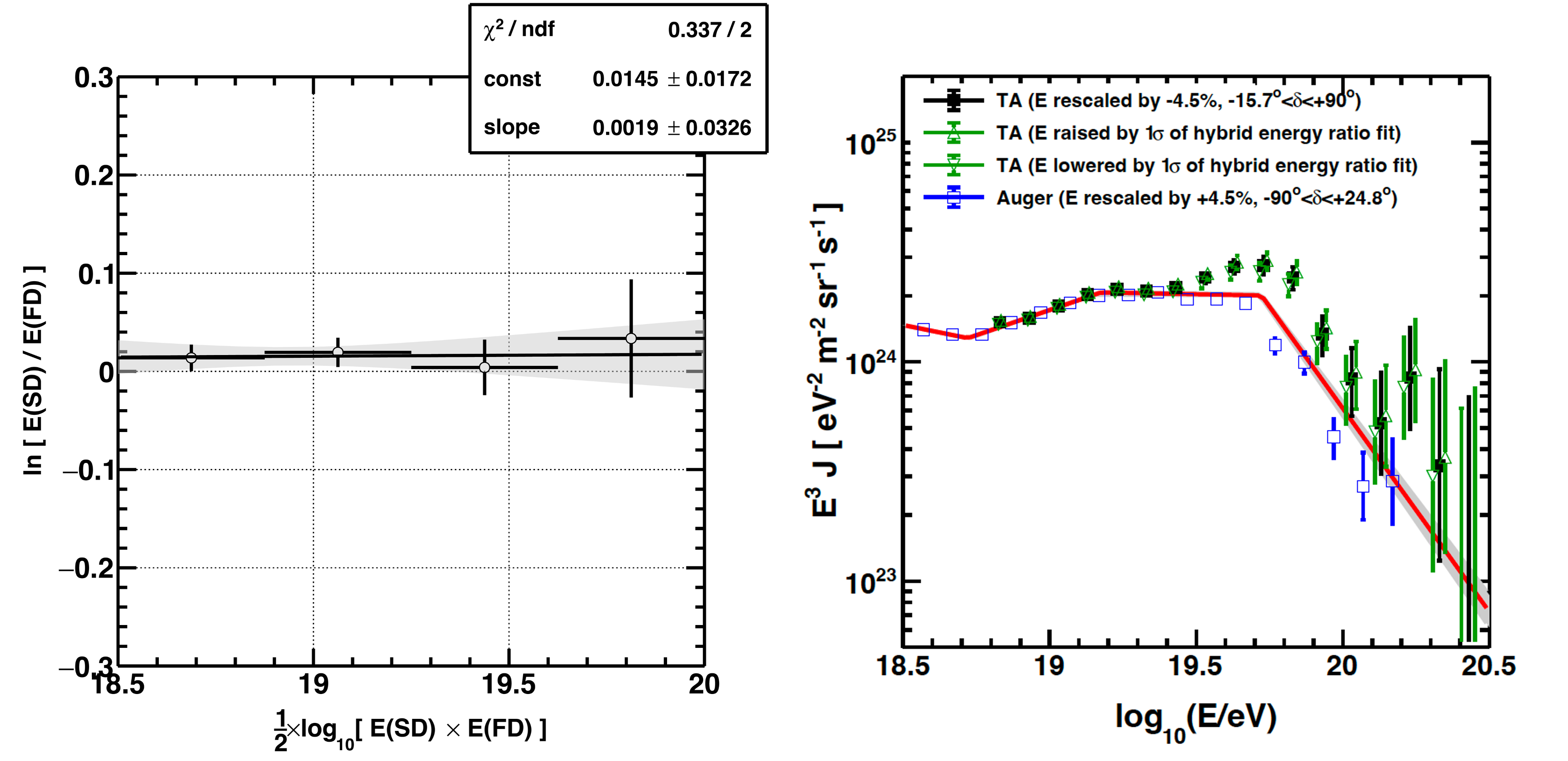}
\caption{\textbf{TA energy-scale check and TA--Auger spectra.} \textbf{Left}: The natural
logarithm of the energy ratio for TA hybrid events as a function of energy. The fitted slope indicates no significant energy dependence in the SD/FD energy ratio. \textbf{Right}: TA and Auger spectra in their full apertures after the relative energy rescaling used in joint comparisons. The TA spectra shifted by $\pm1\sigma$ from the SD/FD energy-ratio fit are also shown with the green arrows. The difference at the highest energies is not removed by this energy-scale uncertainty.}
\label{fig:dec_depend}
\end{figure}

The energy spectrum measured by TA, TALE, and TA$\times$4 spans approximately five orders of magnitude in energy, from the PeV region to beyond $10^{20}$~eV. Over this interval, six spectral features are observed: the knee, the low-energy ankle, the second knee, the ankle, the instep or shoulder feature, and the high-energy cutoff. These features provide important information on the transition from Galactic to extragalactic cosmic rays, the maximum energies of source populations, and propagation effects at the highest energies.

The latest TA SD spectrum uses 16 years of data collected from 11 May 2008 to 10 May 2024~\cite{Kim:2025icrc301}. 
The left panel of Figure~\ref{fig:ta_tax4_spec} shows that the TA SD energy spectrum is well described by a thrice-broken power law. The fit identifies an ankle at $\log_{10}(E_{\rm ankle}/{\rm eV}) = 18.70 \pm 0.01$, a shoulder feature at $\log_{10}(E_{\rm shoulder}/{\rm eV}) = 19.15 \pm 0.08$, and a high-energy cutoff at $\log_{10}(E_{\rm cutoff}/{\rm eV}) = 19.83 \pm 0.03$. The cutoff is observed with a chance probability of $1.6\times10^{-10}$, corresponding to a significance of $6.3\sigma$. The shoulder feature is observed with a chance probability of $1.3\times10^{-7}$, corresponding to $\sim$$5.2\sigma$. The observed shoulder position in the Northern Hemisphere is consistent with the analogous feature reported by the Pierre Auger Observatory (Auger)~\cite{PierreAuger:2025eun} after applying the relative energy-scale treatment of $\sim$9\% used in TA--Auger comparisons.

The stability of the TA SD energy reconstruction has also been tested with the constant-intensity-cut (CIC) method, a data-driven cross-check. The spectrum obtained with the CIC method is consistent with the standard TA reconstruction within about 2\%~\cite{Kim:2025icrc301}. This agreement holds for both the full data set and the declination-binned spectra, providing an internal validation of the zenith-angle attenuation correction and the linearity of the SD energy reconstruction.

The TA$\times$4 SD spectrum has been measured using three years of data collected from October 2019 to September 2022~\cite{Koyama:2025icrc310}. This analysis includes newly reconstructed inclined air-shower events with zenith angles in the range $55^\circ$--$65^\circ$, extending the event sample beyond the previous selection. As shown in the right panel of Figure~\ref{fig:ta_tax4_spec}, the TA$\times$4 SD spectrum is consistent with the TA SD spectrum within the current statistical uncertainties, supporting the use of the expanded array for future high-statistics measurements at the highest energies.

At lower energies, TALE provides spectrum measurements in the PeV--EeV region, as shown in Figure~\ref{fig:tale_spec}. The TALE FD monocular analysis measures the spectrum using fluorescence and Cherenkov light observed by the high-elevation TALE telescopes~\cite{AbuZayyad:2023icrc379}, while the TALE hybrid analysis combines the TALE FD with the TALE SD array~\cite{Oshima:2025icrc350}. The TALE hybrid spectrum exhibits a break at $\log_{10}(E/{\rm eV}) = 17.11 \pm 0.09$ in the energy region associated with the second knee. These TALE measurements connect the lower-energy Galactic cosmic-ray region to the ankle region measured by TA.

The declination dependence of the spectrum provides an important clue to the difference between the northern and southern sky measurements. Considerable effort within the TA--Auger Joint Spectrum Working Group has been devoted to understanding the spectral differences in the common declination band. Many possible sources of systematic uncertainty, including energy-dependent effects, have been investigated extensively, but none has been established as a confirmed explanation~\cite{Verzi:2017hro,specWG_UHECR2018:2019vpk,Deligny:2020gzq,TelescopeArray:2021zox,Tsunesada:2023yhw}. 

To quantify the difference between the two spectra, a simultaneous broken-power-law fit is performed under the null hypothesis that they are drawn from the same parent spectrum~\cite{Kim:2025icrc301}. For the full TA and Auger apertures, the fit gives a probability of $7.5\times10^{-16}$, corresponding to approximately $8.0\sigma$. When the comparison is restricted to the common declination band, with \textit{a priori} established excess regions masked and events below $\delta=-5^\circ$ excluded, where the TA exposure is very small and drops rapidly, the agreement improves substantially, with a probability of $3.8\times10^{-2}$, corresponding to about $1.8\sigma$. 

The possible non-linearity of the TA energy scale has also been evaluated by rescaling the energy according to the $\pm 1\sigma$ uncertainty in $\ln(E_{\rm SD}/E_{\rm FD})$ and repeating the simultaneous TA--Auger fit. Figure~\ref{fig:dec_depend} shows the TA energy-scale check in the left panel and the corresponding simultaneous fit including the TA energy-scale uncertainty in the right panel. This procedure gives a range of significances from approximately $10.0\sigma$ to $5.6\sigma$, with $5.6\sigma$ taken as a conservative value. Thus, the current uncertainty in the energy-scale linearity does not remove the high-energy full-aperture discrepancy between the TA and Auger spectra. These findings suggest that the observed spectral difference arises from a genuine difference in the UHECR flux between the northern and southern skies.

Systematic effects in the TA SD spectrum have been investigated by varying the assumed primary mass composition and hadronic interaction model~\cite{Fujisue:2025icrc259}. As extreme cases, pure-proton and pure-iron Monte Carlo samples based on the QGSJetII-04 model were used to estimate the energy-dependent bias that could arise from the primary-mass assumption and the hadronic model over the full energy range. The study shows that the assumed mass composition and hadronic model can affect the reconstructed spectrum. However, the resulting systematic shift is not large enough to account for the difference between the TA and Auger spectra in the highest-energy region.

\subsection{Mass Composition}
\label{subsec:composition}

\begin{figure}[t]
\centering
\includegraphics[width=0.8\columnwidth]{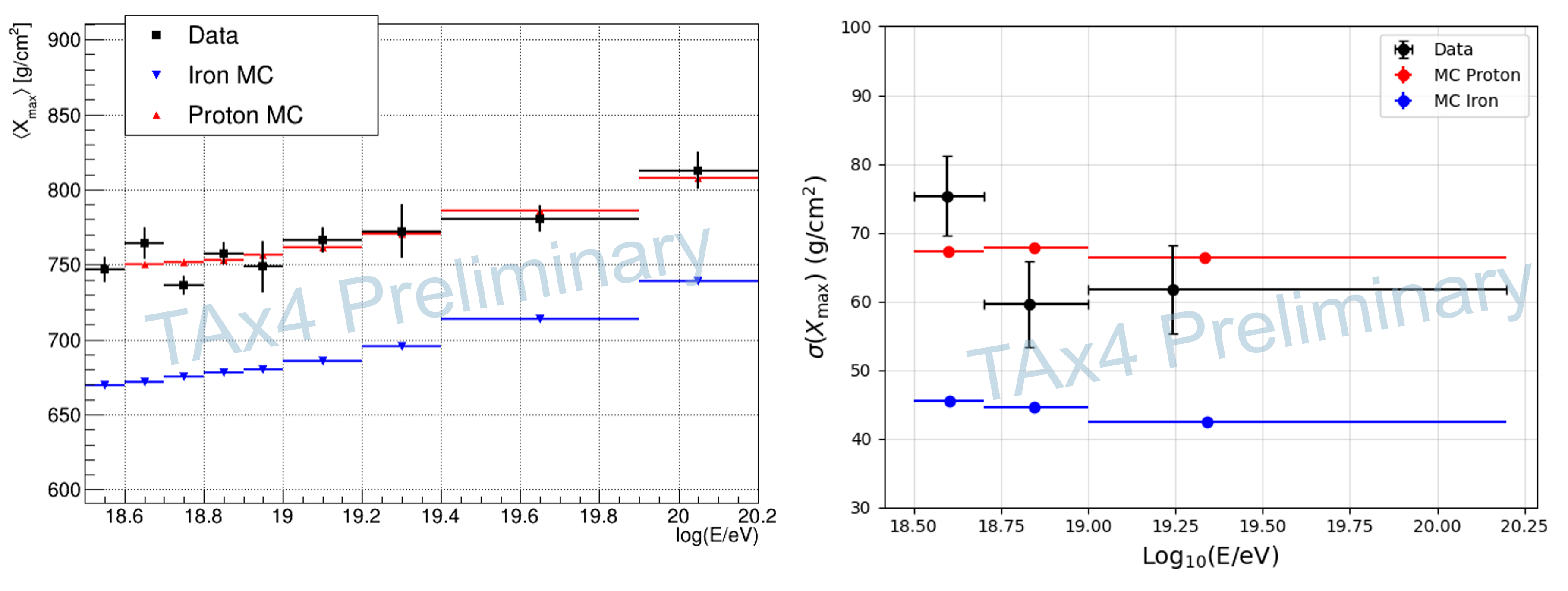}
\caption{\textbf{TA$\times$4 hybrid $X_{\max}$ measurements.} \textbf{Left}: Mean $X_{\max}$ as a function of energy for the TA$\times$4 hybrid data, compared with proton and iron Monte Carlo predictions based on the QGSJetII-03 hadronic interaction model. \textbf{Right}: Corresponding $\sigma(X_{\max})$ values. The data are consistent with Monte Carlo protons. 
}
\label{fig:tax4_xmax}
\end{figure}

\begin{figure}[t]
\centering
\begin{minipage}[t]{0.44\columnwidth}
    \centering
    \includegraphics[width=\columnwidth]{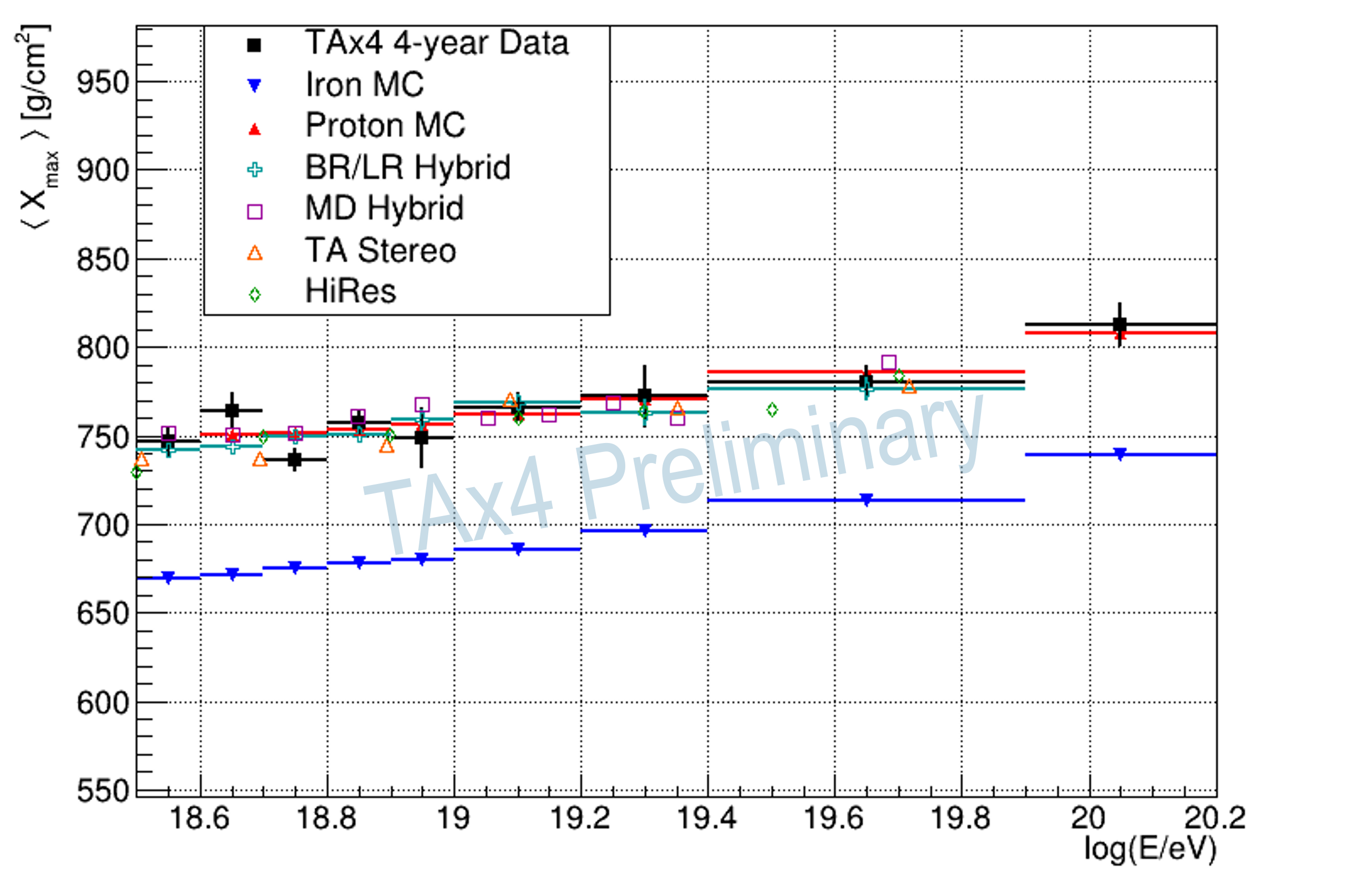}
    \caption{\textbf{$X_{\max}$ measurements in the northern sky.} Various $X_{\max}$ measurements in the northern hemisphere, including HiRes and TA. The northern sky measurements are are consistent with each other.}
    \label{fig:northern_xmax}
\end{minipage}\hfill
\begin{minipage}[t]{0.53\columnwidth}
    \centering
    \includegraphics[width=\columnwidth]
    {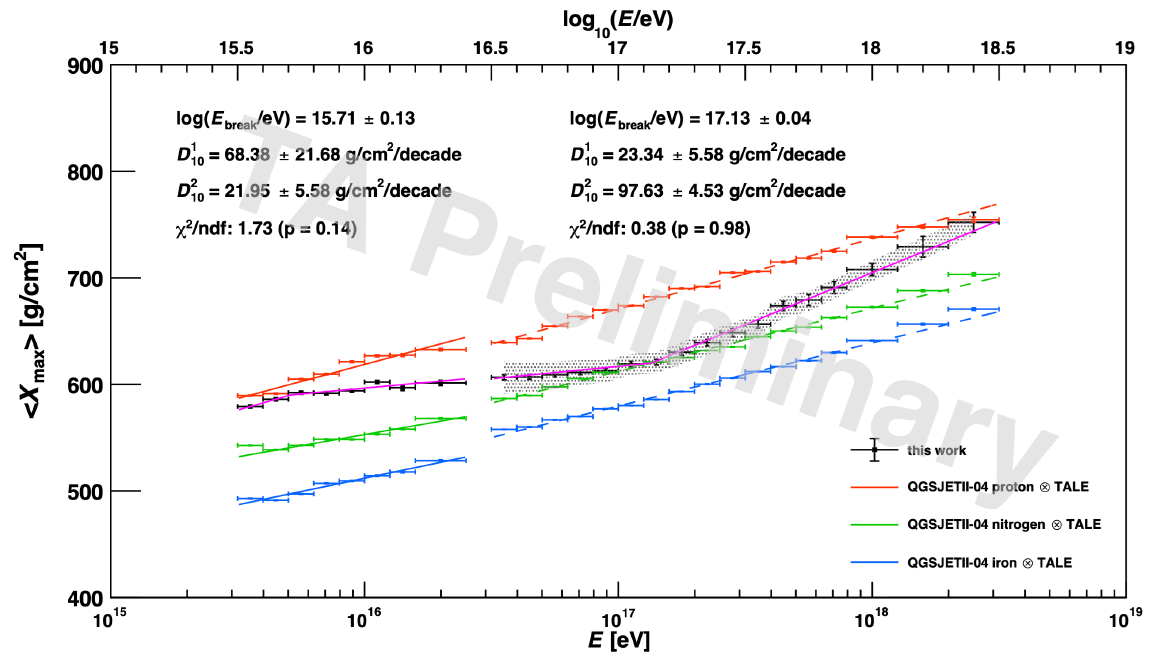}
    \caption{\textbf{TALE and TALE infill $X_{\max}$ results.} Mean $X_{\max}$ versus energy for the TALE and TALE infill hybrid data, compared with proton, nitrogen, and iron Monte Carlo predictions from the QGSJetII-04 hadronic interaction model. The fitted break points and elongation rates are shown on the plot, indicating the composition evolution in the PeV--EeV transition region.}
    \label{fig:tale_xmax}
\end{minipage}
\end{figure}

The mass composition of cosmic rays is primarily studied using the atmospheric depth of shower maximum, $X_{\max}$. On average, light primaries such as protons penetrate more deeply into the atmosphere and have larger $\langle X_{\max}\rangle$ than heavy primaries such as iron. The evolution of $\langle X_{\max}\rangle$ with energy, together with the width of the $X_{\max}$ distribution, therefore provides information on the primary composition. In TA and its extensions, composition is measured mainly with FD and hybrid data, while photon and neutrino searches provide complementary constraints on neutral-particle components.

The TA$\times$4 hybrid composition analysis uses four years of hybrid data collected with the TA$\times$4 FD and TA$\times$4 SD~\cite{Gerber:2025icrc269}. The data are compared with single-element Monte Carlo simulations based on the QGSJetII-03 hadronic interaction model. As shown in the left panel of Figure~\ref{fig:tax4_xmax}, the measured $\langle X_{\max}\rangle$ values are consistent with the proton Monte Carlo prediction in both absolute scale and elongation rate. The right panel shows the $\sigma(X_{\max})$ values, which are less model-dependent but require large statistics for a reliable measurement. These are also compatible with the proton prediction within the current uncertainties. Figure~\ref{fig:northern_xmax} compares $\langle X_{\max}\rangle$ measurements in the Northern Hemisphere, including results from HiRes and TA~\cite{HiRes:2009fiy,TelescopeArray:2014sfa,TelescopeArray:2018xyi,Bergman:2019pj}, and shows that they are mutually consistent within uncertainties.

\begin{figure}[tbp]
\centering
\includegraphics[width=\columnwidth]{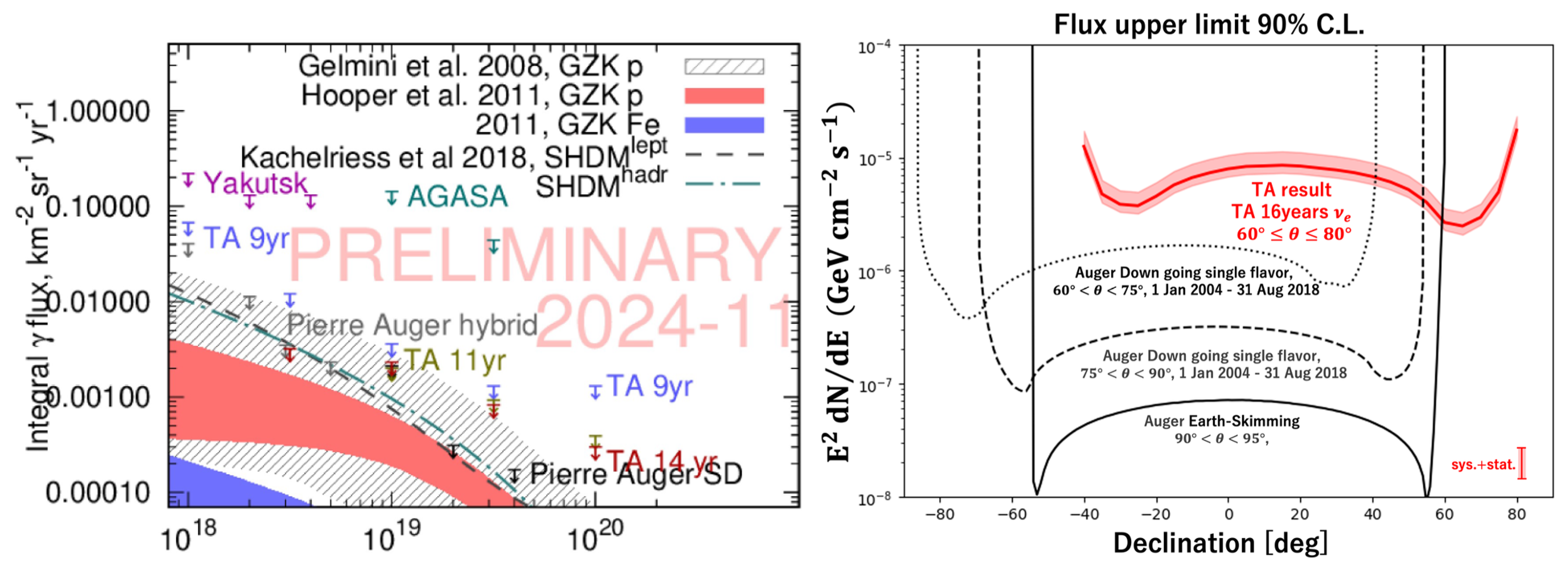}
\caption{\textbf{TA searches for UHE photons and neutrinos.} \textbf{Left}: TA integral photon flux upper limits, compared with previous measurements and selected model predictions. The TA limits improve constraints on UHE photon fluxes over a broad energy range. \textbf{Right}: 90\% C.L. upper limits on UHE neutrinos are obtained with 16 years of TA SD data as a function of declination, compared with limits from Auger. The comparison highlights the complementary sky coverage of the two experiments.}
\label{fig:photon_neutrino}
\end{figure}

The TA$\times$4 hybrid data set is especially important because it extends hybrid composition measurements into the highest energy region. The current analysis shows reasonable agreement between data and Monte Carlo simulations, and includes events above 100~EeV~\cite{Gerber:2025icrc269}. Continued operation of TA$\times$4 will improve the statistical precision of $\langle X_{\max}\rangle$ and $\sigma(X_{\max})$ in the region above $10^{19}$~eV, where composition is critical for interpreting anisotropy searches and the cutoff.

At lower energies, the TALE FD monocular data set provides composition information from approximately $10^{15}$ to $10^{18}$~eV. The eight-year TALE FD monocular analysis, using data from June 2014 to August 2022, finds a break in the elongation rate at $E \simeq 10^{17.22 \pm 0.04}~{\rm eV}$ near the second knee~\cite{AbuZayyad:2023icrc379}. Fits of the measured $X_{\max}$ distributions indicate a light--heavy--light composition evolution: the composition becomes heavier approaching the second-knee region and then becomes lighter again toward EeV energies. This behavior is consistent with a transition between Galactic and extragalactic cosmic ray components.

The TALE and TALE-infill hybrid measurements provide an additional hybrid view of the same transition region~\cite{Fujita:2025qfo,TelescopeArray:2026rdu}. Breaks in $\langle X_{\max}\rangle$, shown in Figure~\ref{fig:tale_xmax}, coincide with the knee and second knee, and fits to the $X_{\max}$ distributions show a transition from light to heavier composition, peaking near $E \simeq 10^{17.13}~{\rm eV}$. At higher energies the composition becomes lighter again. Further studies including helium nuclei and updated hadronic interaction models, such as EPOS-LHC-R and QGSJetIII, are ongoing.

UHE photon searches provide an additional probe of the primary composition and of possible cosmogenic or exotic components. The TA SD photon analysis uses a neural-network classifier trained on photon- and proton-induced Monte Carlo air showers and further fine-tuned with data events identified with high confidence as non-photon-like~\cite{Abbasi2019_TA_DiffusePhotonFlux,Rubtsov:2024uhecrPhotons}. The left panel of Figure~\ref{fig:photon_neutrino} shows the resulting upper limits on the diffuse UHE photon flux obtained with 14 years of TA SD data. 

The TA SD has also been used to search for UHE neutrinos through deeply developing inclined air showers with $60^\circ < \theta < 80^\circ$. No neutrino candidate was found in 16 years of TA SD data, and 90\% confidence level upper limits on the diffuse UHE neutrino flux were derived~\cite{Takahashi:2025icrc1192}, as shown in the right panel of Fig.~\ref{fig:photon_neutrino}. This analysis demonstrates the capability of the TA SD to constrain UHE neutrino fluxes in the northern sky and provides a foundation for future diffuse, point-source, and transient searches.


\subsection{Anisotropies}
\label{subsec:anisotropy}

\begin{figure}[t]
\centering
\includegraphics[width=1\columnwidth]{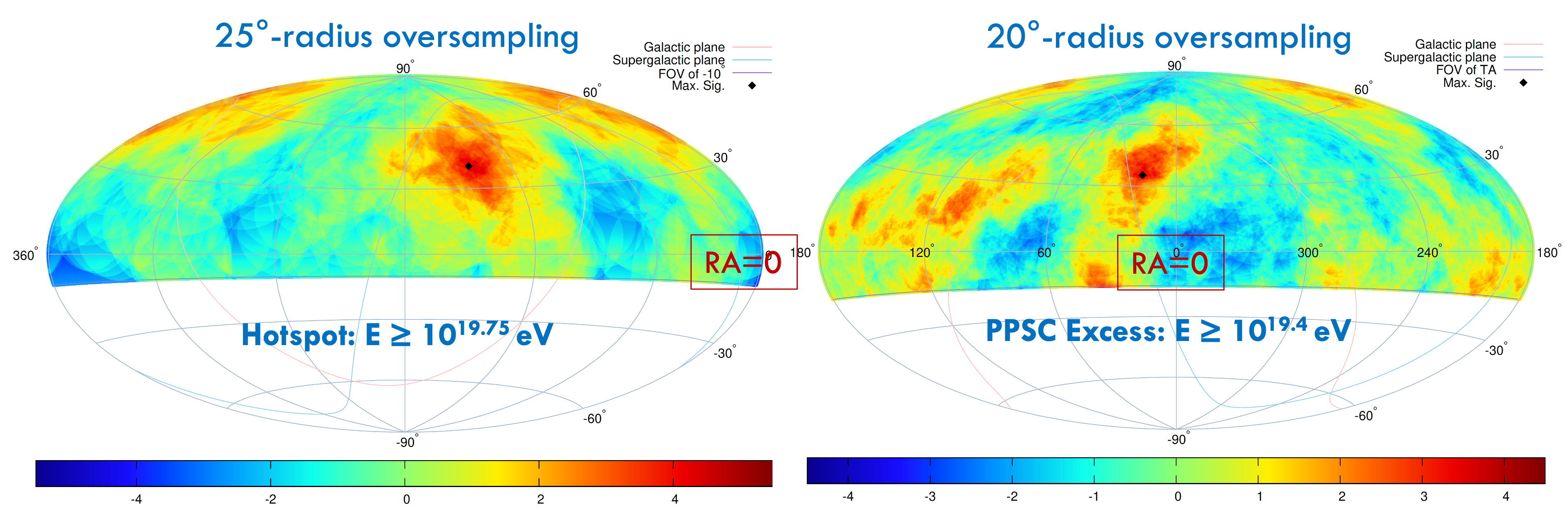}
\caption{\textbf{Intermediate-scale anisotropy skymaps.} \textbf{Left}: TA Hotspot Li-Ma significance map obtained with $25^\circ$-radius oversampling for events with $E \geq 10^{19.75}$~eV. \textbf{Right}: PPSC excess Li-Ma significance map obtained with $20^\circ$-radius oversampling for events with $E \geq 10^{19.4}$~eV. The diamond indicates the maximum significance in each map. The Galactic plane, Supergalactic plane, and the field of view are also shown. }
\label{fig:intermediate_anisotropies}
\end{figure}

\begin{wrapfigure}{r}{0.5\textwidth}
\centering
\includegraphics[width=0.5\columnwidth]
{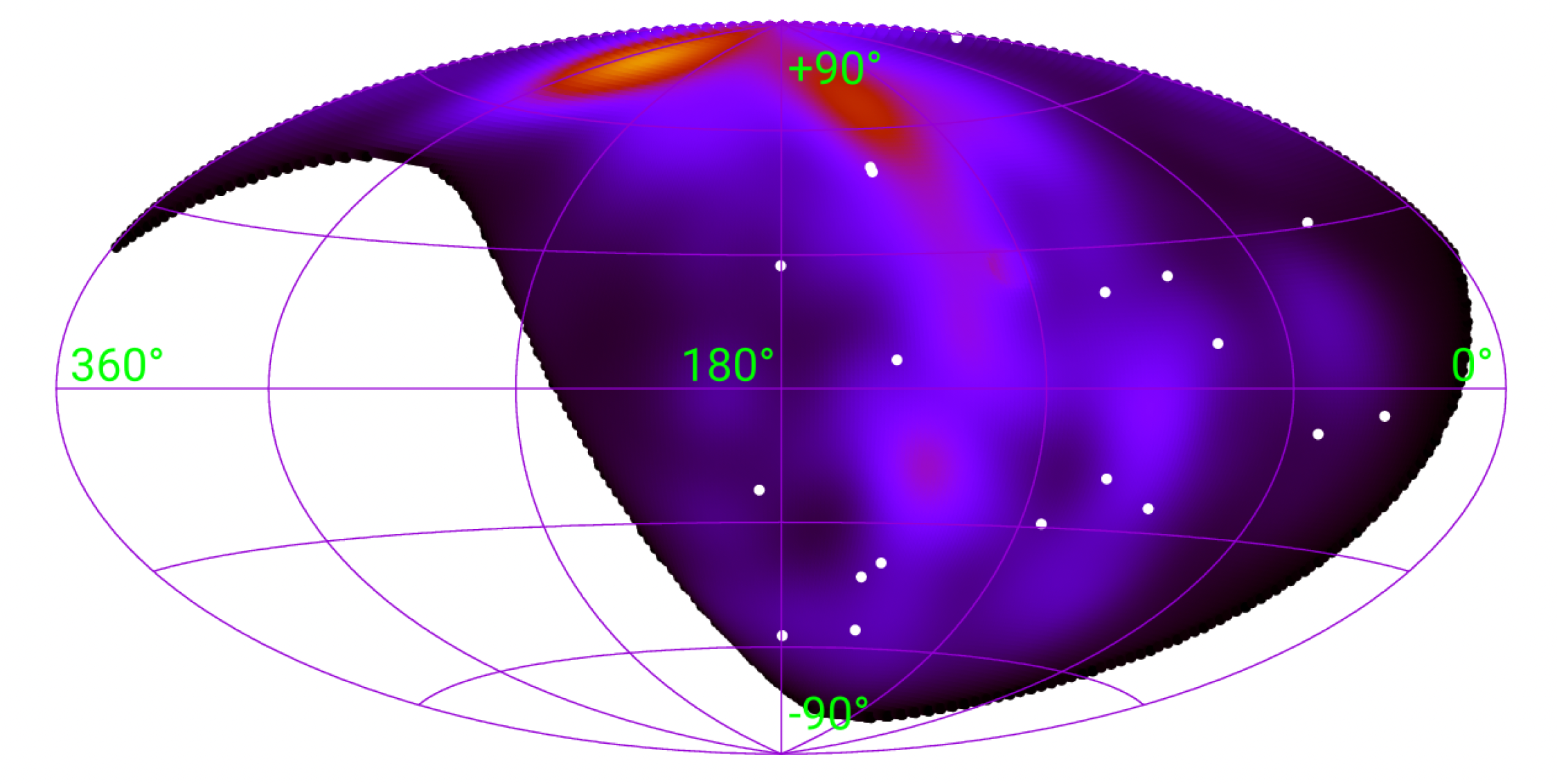}
\caption{\textbf{LSS flux map together with events above 100~EeV.} Expected UHECR flux from the LSS model assuming pure protons, with a $10^\circ$ smearing angle, is shown in Hammer projection in Galactic coordinates. White filled circles show the 19 observed events with $E > 100$~EeV. The observed event distribution is not compatible with this LSS model prediction.}
\label{fig:lss_anisotropy}
\end{wrapfigure}

Anisotropy in the arrival direction distribution is a key observable for identifying UHECR sources. At the highest energies, the reduced energy loss horizon may enhance the visibility of source structures. However, the interpretation of anisotropy depends strongly on the primary mass composition and on magnetic deflections in Galactic and extragalactic magnetic fields. The TA dataset provides the largest Northern Hemisphere exposure for such studies.

Intermediate-scale anisotropies have been studied with TA SD data using an oversampling method, in which the number of UHECR events inside a fixed-radius circle is counted across the sky and compared with the isotropic expectation using the Li--Ma significance. The original TA hotspot was reported using the first five years of TA SD data above 57~EeV, with an excess near the Ursa Major region~\cite{Abbasi:2014hotspot}. The updated analysis uses 16 years of TA SD data containing 228 events with energies greater than 57~EeV. Using a $25^\circ$-radius oversampling window, the maximum local significance is $4.9\sigma$ at $(\alpha,\delta) = (144.0^\circ,40.5^\circ)$, with a global significance of $2.9\sigma$~\cite{Kim:2025icrc302}, as shown in Figure~\ref{fig:intermediate_anisotropies} (left). The excess therefore persists near the direction of the Ursa Major constellation, although additional exposure is required to determine whether it represents a genuine astrophysical feature.

A second intermediate-scale excess is observed at somewhat lower energies in the direction of the Perseus--Pisces supercluster (PPSC), as shown in Figure~\ref{fig:intermediate_anisotropies} (right). The updated analysis considers energy thresholds of $E>10^{19.4}$~eV, $E>10^{19.5}$~eV, and $E>10^{19.6}$~eV, corresponding to 1186, 767, and 464 events, respectively~\cite{Kim:2025icrc302}. Using a $20^\circ$-radius oversampling window, the maximum local significances are $3.7\sigma$, $3.9\sigma$, and $3.7\sigma$ for these three thresholds. The chance probabilities for obtaining an equal or larger excess near the PPSC direction correspond to $3.1\sigma$, $3.2\sigma$, and $3.0\sigma$, respectively. These results suggest a possible association between UHECR arrival directions and a source or sources in the direction of the PPSC.

The correlation of the highest-energy TA events with nearby large-scale structure (LSS) has also been studied using source distributions based on the 2MASS Redshift Survey catalog within 250~Mpc~\cite{Abbasi:2024prlLSS,Abbasi:2024prdLSS}. In this modeling, UHECR propagation is simulated, Galactic magnetic field models are included, and extragalactic magnetic fields are represented by angular smearing. The arrival directions of 19 TA events with $E>100$~EeV were compared with the predicted LSS flux maps, as shown in Figure~\ref{fig:lss_anisotropy}. The observed data are compatible with isotropy. For the LSS model to remain compatible with the data at the $2\sigma$ level, the interpretation requires either a very heavy primary composition, heavier than iron, or, in the case of proton or helium primaries, strong extragalactic magnetic fields with strengths above approximately 20~nG for a coherence length of 1~Mpc.

\begin{figure}[t]
\centering
\includegraphics[width=0.8\columnwidth]{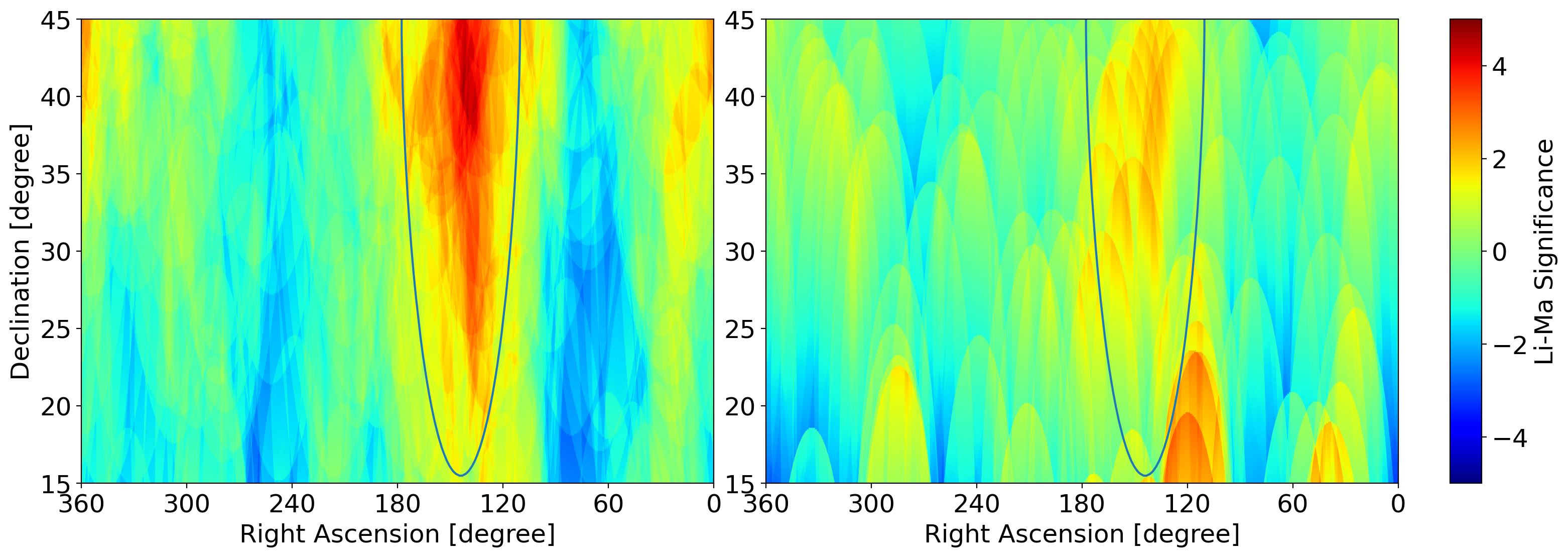}
\caption{\textbf{Comparison of the TA hotspot using declination weights in equatorial coordinates.} \textbf{Left}: TA Hotspot observed by TA. \textbf{Right}: TA Hotspot as it would appear in the Auger detector after applying declination weights to the TA events. The blue solid circles mark the $25^\circ$ window centered on the hotspot.}
\label{fig:weighted_skymap}
\end{figure}

The anisotropy results suggest that the Northern sky may contain intermediate-scale structures at the highest energies. Although the Auger has searched for excesses in the directions of the TA Hotspot and the PPSC excess, no statistically significant signals were found~\cite{PierreAuger:2024hrj}. We demonstrated that this is consistent with Auger's limited exposure at the high declinations where these excesses appear~\cite{Kim:2025icrc302}. To illustrate this effect, the TA hotspot was weighted by the ratio of Auger to TA exposure as a function of declination. Figure~\ref{fig:weighted_skymap} shows the TA Hotspot for reference (left) and the weighted event distribution as it would appear to Auger (right). In the hotspot region, defined by $15^\circ < \delta < 45^\circ$, the local Li--Ma significance at $(144.0^\circ, 40.5^\circ)$ is reduced from about $5\sigma$ in TA to about $2\sigma$ after weighting. Because Auger exposure decreases rapidly at high declinations, it is much less sensitive to northern-sky anisotropies at the present level of exposure.

\section{Summary}
\label{sec:summary}

The Telescope Array experiment and its extensions provide a broad Northern Hemisphere observation for measuring cosmic rays from the PeV scale to beyond $10^{20}$~eV. The combined TALE, TALE infill, TA, and TA$\times$4 detector systems use hybrid observations to connect the energy spectrum, mass composition, and arrival direction information across this wide energy range.

Recent spectrum measurements cover approximately five orders of magnitude in energy and reveal six major spectral features: the knee, low-energy ankle, second knee, ankle, instep or shoulder feature, and high-energy cutoff. In the updated 8-year TALE FD monocular spectrum, the low-energy ankle is observed at $\log_{10}(E/{\rm eV}) = 16.27 \pm 0.01$ and the second knee at $\log_{10}(E/{\rm eV}) = 17.04 \pm 0.02$. In the latest 16-year TA SD spectrum, the ankle is confirmed at $\log_{10}(E/{\rm eV}) = 18.70 \pm 0.01$, the shoulder at $\log_{10}(E/{\rm eV}) = 19.15 \pm 0.08$, and the cutoff at $\log_{10}(E/{\rm eV}) = 19.83 \pm 0.03$. The TA$\times$4 SD spectrum, including newly reconstructed inclined events, is consistent with the TA SD spectrum and demonstrates the readiness of the expanded array for high-statistics measurements at the highest energies.

Mass composition measurements show a light--heavy--light evolution between $10^{15}$ and $10^{18}$~eV in TALE FD monocular and hybrid analyses, with changes in the elongation rate associated with the knee and second knee. Above $10^{18}$~eV, TA and TA$\times$4 hybrid data remain consistent with a light and approximately steady composition, although statistics above $10^{19}$~eV are still limited. Searches for UHE photons and neutrinos using TA SD data provide complementary constraints on neutral-particle components.

Arrival direction studies continue to show intriguing intermediate-scale structures in the Northern sky. The TA Hotspot persists near the direction of the Ursa Major constellation, and a second excess is observed at somewhat lower energies in the direction of the Perseus--Pisces supercluster. At the same time, comparisons of events above 100~EeV with nearby large scale structure source models may indicate that either a heavy composition or strong magnetic deflections are required for consistency with those models. Completing TA$\times$4 and accumulating its exposure will be crucial for improving composition and anisotropy measurements and for clarifying the origin of UHECRs.

\section*{Acknowledgments}
The Telescope Array experiment is supported by the Japan Society for
the Promotion of Science(JSPS) through
Grants-in-Aid
for Priority Area
431,
for Specially Promoted Research
JP21000002,
for Scientific Research (S)
JP19104006,
for Specially Promoted Research
JP15H05693,
for Scientific Research (S)
JP19H05607,
for Scientific Research (S)
JP15H05741,
for Science Research (A)
JP18H03705,
for Young Scientists (A)
JPH26707011,
for Transformative Research Areas (A)
JP25H01294,
for International Collaborative Research
24KK0064,
and for Fostering Joint International Research (B)
JP19KK0074,
by the joint research program of the Institute for Cosmic Ray Research (ICRR), The University of Tokyo;
by the Pioneering Program of RIKEN for the Evolution of Matter in the Universe (r-EMU);
by the U.S. National Science Foundation awards
PHY-1806797, PHY-2012934, PHY-2112904, PHY-2209583, PHY-2209584, and PHY-2310163, as well as AGS-1613260, AGS-1844306, and AGS-2112709;
by the National Research Foundation of Korea
(2017K1A4A3015188, 2020R1A2C1008230, and RS-2025-00556637) ;
by the Ministry of Science and Higher Education of the Russian Federation under the contract 075-15-2024-541, IISN project No. 4.4501.18, by the Belgian Science Policy under IUAP VII/37 (ULB), by National Science Centre in Poland grant 2020/37/B/ST9/01821, by the European Union and Czech Ministry of Education, Youth and Sports through the FORTE project No. CZ.02.01.01/00/22\_008/0004632, and by the Simons Foundation (MP-SCMPS-00001470, NG). This work was partially supported by the grants of the joint research program of the Institute for Space-Earth Environmental Research, Nagoya University and Inter-University Research Program of the Institute for Cosmic Ray Research of University of Tokyo. The foundations of Dr. Ezekiel R. and Edna Wattis Dumke, Willard L. Eccles, and George S. and Dolores Dor\'e Eccles all helped with generous donations. The State of Utah supported the project through its Economic Development Board, and the University of Utah through the Office of the Vice President for Research. The experimental site became available through the cooperation of the Utah School and Institutional Trust Lands Administration (SITLA), U.S. Bureau of Land Management (BLM), and the U.S. Air Force. We appreciate the assistance of the State of Utah and Fillmore offices of the BLM in crafting the Plan of Development for the site. We thank Patrick A.~Shea who assisted the collaboration with much valuable advice and provided support for the collaboration's efforts. The people and the officials of Millard County, Utah have been a source of steadfast and warm support for our work which we greatly appreciate. We are indebted to the Millard County Road Department for their efforts to maintain and clear the roads which get us to our sites. We gratefully acknowledge the contribution from the technical staffs of our home institutions. An allocation of computing resources from the Center for High Performance Computing at the University of Utah as well as the Academia Sinica Grid Computing Center (ASGC) is gratefully acknowledged.

\section*{References}
\small

\begin{thebibliography}{99}
\bibitem{TelescopeArray:2012uws}
T.~Abu-Zayyad {\it et al.}, \Journal{\NIMA}{689}{87--97}{2013}.

\bibitem{TelescopeArray:2012vhh}
T.~Abu-Zayyad {\it et al.}, \Journal{\ASP}{39--40}{109--119}{2012}.

\bibitem{Tokuno:2012mi}
H.~Tokuno {\it et al.}, \Journal{\NIMA}{676}{54--65}{2012}.

\bibitem{TelescopeArray:2018bya}
R.~U. Abbasi {\it et al.}, \Journal{\APJ}{865}{74}{2018}.

\bibitem{Komae:2023fsb}
Ichiro Komae {\it et al.},, {\em PoS(ICRC2023)}405.

\bibitem{Kawachi:2025bbk}
Yusuke Kawachi {\it et al.},, {\em PoS(UHECR2024)}096.

\bibitem{TelescopeArray:2021dri}
R.~U. Abbasi {\it et al.}, \Journal{\NIMA}{1019}{165726}{2021}.

\bibitem{Abbasi2020_ProtonAirCrossSection_MD}
R.~U. Abbasi {\it et al.}, \Journal{\PRD}{92}{032007}{2015}.

\bibitem{Abbasi2020_ProtonAirCrossSection_BRM_LR}
R.~U. Abbasi {\it et al.}, \Journal{\PRD}{102}{062004}{2020}.

\bibitem{Abbasi2018_TA_MuonStudy}
R.~U. Abbasi {\it et al.}, \Journal{\PRD}{98}{022002}{2018}.

\bibitem{Abbasi2019_TA_DiffusePhotonFlux}
R.~U. Abbasi {\it et al.}, \Journal{\ASP}{110}{8--14}{2019}.

\bibitem{Abbasi2020_TA_NeutrinoSearch}
R.~U. Abbasi {\it et al.}, {\em J. Exp. Theor. Phys.} \textbf{131}, 255--264 (2020).

\bibitem{Shinozaki2023_DIMS_TA}
K.~Shinozaki {\it et al.}, {\em PoS(ICRC2023)}1390.

\bibitem{Belz2020_TA_DownwardTGFOrigin}
J.~W. Belz {\it et al.}, {\em J. Geophys. Res. Atmos.} \textbf{125}, e2019JD031940 (2020).

\bibitem{JEMEUSO2015_EUSOTA_GroundTests}
JEM-EUSO Collaboration, {\em Exp. Astron.} \textbf{40}, 301--314 (2015).

\bibitem{Malacari2020_FAST_Prototypes}
M.~Malacari {\it et al.}, \Journal{\ASP}{119}{102430}{2020}.

\bibitem{Tameda2023_CRAFFT}
Y.~Tameda {\it et al.}, \Journal{\EPJ}{283}{06011}{2023}.

\bibitem{Mayotte2023_AugerAtTA}
S.~Mayotte {\it et al.}, {\em PoS(ICRC2023)}368.

\bibitem{Kim:2025icrc301}
J. Kim {\it et al.}, {\em PoS(ICRC2025)}301.

\bibitem{PierreAuger:2025eun}
Adila Abdul~Halim {\it et al.}, \Journal{\PRL}{135}{241002}{2025}.

\bibitem{Koyama:2025icrc310}
Chisato Koyama {\it et al.}, {\em PoS(ICRC2025)}310.

\bibitem{AbuZayyad:2023icrc379}
T.~AbuZayyad {\it et al.}, {\em PoS(ICRC2023)}379.

\bibitem{Oshima:2025icrc350}
H.~Oshima {\it et al.}, {\em PoS(ICRC2025)}350.

\bibitem{Verzi:2017hro}
V.~Verzi {\it et al.}, \Journal{\PTEP}{2017}{12A103}{2017}.

\bibitem{specWG_UHECR2018:2019vpk}
D.~Ivanov {\it et al.}, \Journal{\EPJ}{210}{01002}{2019}.

\bibitem{Deligny:2020gzq}
O.~Deligny {\it et al.},, {\em PoS(ICRC2019)}234.

\bibitem{TelescopeArray:2021zox}
Y.~Tsunesada {\it et al.}, {\em PoS(ICRC2021)}337.

\bibitem{Tsunesada:2023yhw}
Y.~Tsunesada {\it et al.}, {\em PoS(ICRC2023)}406.

\bibitem{Fujisue:2025icrc259}
K.~Fujisue {\it et al.}, {\em PoS(ICRC2025)}259.

\bibitem{Gerber:2025icrc269}
Z.~Gerber {\it et al.}, {\em PoS(ICRC2025)}269.

\bibitem{HiRes:2009fiy}
R.~U. Abbasi {\it et al.}, \Journal{\PRL}{104}{161101}{2010}.

\bibitem{TelescopeArray:2014sfa}
R.~U. Abbasi {\it et al.}, \Journal{\ASP}{64}{49--62}{2015}.

\bibitem{TelescopeArray:2018xyi}
R.~U. Abbasi {\it et al.}, \Journal{\APJ}{858}{76}{2018}.

\bibitem{Bergman:2019pj}
D.~Bergman {\it et al.}, {\em PoS(ICRC2019)}191.

\bibitem{Fujita:2025qfo}
K.~Fujita {\it et al.}, {\em PoS(ICRC2025)}261.

\bibitem{TelescopeArray:2026rdu}
R.~U. Abbasi {\it et al.}, \Journal{\PRD}{113}{062003}{2026}.

\bibitem{Rubtsov:2024uhecrPhotons}
G.~Rubtsov {\it et al.}, Talk presented at the 7th International Symposium on Ultra High Energy Cosmic Rays (UHECR2024), Malarg{\"u}e, Mendoza, Argentina, 17--21 November 2024.

\bibitem{Takahashi:2025icrc1192}
K.~Takahashi {\it et al.}, {\em PoS(ICRC2025)}1192.

\bibitem{Abbasi:2014hotspot}
R.~U. Abbasi {\it et al.}, \Journal{\APJL}{790}{L21}{2014}.

\bibitem{Kim:2025icrc302}
J. Kim {\it et al.}, {\em PoS(ICRC2025)}302.

\bibitem{Abbasi:2024prlLSS}
R.~U. Abbasi {\it et al.}, \Journal{\PRL}{133}{041001}{2024}.

\bibitem{Abbasi:2024prdLSS}
R.~U. Abbasi {\it et al.}, \Journal{\PRD}{110}{022006}{2024}.

\bibitem{PierreAuger:2024hrj}
Adila Abdul~Halim {\it et al.}, \Journal{\APJ}{984}{123}{2025}.

\end{thebibliography}


\end{document}